# Anomalous Low-Frequency Dielectric Response of Clay-Water System

Dean Korošak[a], Janja Kramer[a], Renata Jecl[a], Bruno Cvikl[a,b] and Miran Veselič[c]
[a]Chair for Applied Physics, Faculty of Civil Engineering,
University of Maribor, Smetanova 17, 2000 Maribor, and
[b]"J. Stefan" Institute, Jamova 19, Ljubljana
[c]Agency for Radwaste Management, Parmova 53, Si-1000 Ljubljana
dean.korosak@uni-mb.si

## Abstract

In this work we give the analysis of the low frequency (100 Hz – 100 MHz) dielectric response of selected kaolinitic clays at different water content ranging from dry, over plastic to liquid limit. For all samples investigated, depending on the moisture content, the real part of the dielectric function reflects the abnormal behaviour within the certain frequency region where the negative values of real part of dielectric response have been observed. The anomalies are explained within a generalized conductivity model based on the clay-water electrolyte ions free motion in pore water and their restricted motion near the clay particle surface. The results indicate that the dynamics of the ions in moist clay is in part governed by anomalous diffusion.

## 1. Introduction

Safety and environmental issuses are crucial for the present efforts for determination of radioactive waste repository site in Republic of Slovenia [1]. The soil in the vicinity of the repository site can become contaminated through transport of moist in soil and migration of radionuclides through porous soil [2], so the knowledge and application of complementary experimental methods of contamination detection is of paramount importance [3, 4].

Low and intermedium radioactive waste (LILW) are deposited in surface or underground repositories. The release of radionuclides from the repository into biosphere is prevented using artificial barriers such as concrete containers or concrete walls, and natural barriers such as soil, clay and rock surrounding the repository. Monitoring of radionuclide concentrations in the repository surrounding soil is one of the key safety issues. Often, geophysical methods such as GPR (ground penetrating radar) methods are used in monitoring radionuclide concentration in the vicinity of the repository. For more accurate determination of the contamination the geophysical methods need to be calibrated with in-situ measurements of the soil samples.

The dielectric spectroscopy of selected soil samples offers a possibility for monitoring the contamination of the soil in the vicinity of the respository. However, the complex dielectric characteristics of moist soil [5-8] which includes several relaxation mechanisms and often anomalous features of the underlying transport mechanisms does not allow simple analysis and correlation between the dielectric properties and degree of contamination [9, 10].

The microscopic properties of diffusion transport of ions in clayey soils has recently become an important issue in electromigration studies of radionuclides in moist soil for soil decontamination and engineering nuclear waste repository natural barriers [2, 21-25]. The importance of the effective diffusion constant was emphasizes in ref. [2], where the diffusion coefficients for $^{85}$Sr, $^{131}$I and HTO were obtained, while the dependence of the diffusion coefficient on the distance from the clay-water interface was studied in ref. [21] showing the increase in diffusion coefficient near the interface with respect to its value in the bulk as much as ten times. This leads to the suggestion that by detailed analysis of the dielectric response from contaminated and noncontaminated samples it is possible to obtain additional information on transport properties of contaminants in moist clayey soils.

The aim of this work is to contribute to better understanding of the key physical processes involved in the dielectric response of moist soils and to indentify the parameters that determine the linear response of the clean and contaminated moist soil. Here we concentrate on the very simple non-contaminated system consisting of selected kaolinitic clay and water, but which already exhibits surprisingly complex behaviour.

## 2. Experimental details and results of measurements

The electrical characteristics of the clay-water system were measured using the low frequency impedance analyzer at room temperature. The admittance of the sample placed in the measuring cell between two planparallel electrodes (area $S$=5,5 cm$^2$, distance $L$=4,5 – 5 mm) was determined from the linear response of the sample to the small oscillating bias on the electrodes of magnitude 10 mV. The real (conductance) and imaginary part (capacitance) of the admittance were measured in the frequency interval 100 Hz to 100 MHz.

The clay sample used was kaolinite intended for basic research purposes so the mineral composition and structural formula of the sample were known. The plastic and liquid limit of the sample were 25,9 % and 40,1 % water content respectively. The specific surface of the clay sample was 10 m$^2$/g and specific gravity 2,6.

The frequency dependence of the conductance and capacitance was determined first for dry sample, and then for moist samples, with moisture content ranging from 36 % to 56 %. Moist samples were obtained with addition of distilled water to the clay.

The results of the measured conductance of the dry and of the moist samples with lowest and highest moisture content are shown on figures 1 and 2. The common feature of all capacitance spectra are the increasing capacitance with diminishing frequency. Additionally, moist samples exhibit negative capacitance effect in the bounded frequency region as shown in figure 3. We notice that the conductance spectra of the dry and moist samples exhibit completely different dependence on the frequency (figs. 1 and 2).

## 3. Theoretical model

The water in soils is classified into adsorption, capillary, and free water [5]. We envisage that the motion of ions associated with the first two types is controlled by the surface of the soil particle, while this boundary has no impact on the water molecules and ions associated with the free water. The response of the clay-water system to oscillating electrical field then consists of three parts, so the generalized conductivity of clay-water system is the sum of the bulk conductivity from ions in pore electrolyte, diffusion conductivity from ions in diffusive layer near the clay particle surface, and conductivity from ions in free water. The model for the real part of the frequency dependent conductivity which matches the measured linear response spectra of our clay-water samples (figs. 1-3) is:

$$\text{Re}[\sigma(\omega)] = \frac{\sigma_b}{1+(\omega\tau_b)^2} + \sigma_{s,\alpha}(\omega\tau_s)^\alpha + \sigma_f. \tag{1}$$

Here, $\alpha$ describes type of diffusion and is in the range $-1 < \alpha < 1$, $\tau$ is the typical relaxation time of the processes, and $\sigma_b, \sigma_{s,\alpha}, \sigma_f$ are the porse electrolyte, surface layer and free water (frequency independent in the frequency interval in our experiments) conductivities, respectively.

Let us explain briefly the origin of particular terms of eq. (1). Generalized conductivity, given by Kubo formula [12], is determined with the velocity autocorrelation function of the particles $C_v(t) = \langle v(t)v(0) \rangle$:

$$\sigma(\omega) = \frac{q^2 n}{kT} \int_0^\infty \langle v(t)v(0) \rangle \exp(i\omega t) dt. \tag{2}$$

It has been demonstrated recently [13, 14] that the velocity autocorrelation function for a anomalously diffusing particles is:

$$C_v(t) = \frac{kT}{m} E_{2-p}\left(-(t/\tau)^{2-p}\right), \tag{3}$$

where $E_p(t) = \sum_{n=0}^\infty t^n / \Gamma(pn+1)$ is the Mittag-Lefler function which reduces to exponential function when $p=1$ corresponding to normal diffusion. The asymptotic behavior at long times of the function (3) is an inverse power-law: $C_v(t) \propto t^{p-2}$. The autocorrelation function (3) describes subdiffusion for $0<p<1$ and superdiffusion for $1<p<2$ [13, 14]. The first term of eq. (1) representing the conductivity of the particles in pore electrolyte is obtained from generalized conductivity for $p=1$, while the second term representing the conductivity of the particles in the diffusive layer near the clay particle exhibits anomalous diffusion dynamics with $\alpha = 1 - p$.

The dynamics of ions in the free pore electrolyte and near the clay particle surface is governed by different timescales. In the diffusive surface layer it is characterized with the time constant [17]:

$$\tau_s = \frac{a^2}{2 D_{eff}}, \tag{4}$$

describing dynamics along the clay particle surface. Here, $a$ is the radius of the clay particle, and $D_{eff}$ is the effective diffusion constant. The motion of the ions in the pore electrolyte is characterized with the following time constant based on interface polarization effect in heterogenous dielectrics [18, 19]:

$$\tau_b = \frac{\varepsilon_0 (2\varepsilon_w + \varepsilon_{clay})}{\sigma_{eff}}, \quad (5)$$

describing the dynamics of the ions moving perpendicular to the clay particle surface, where $\varepsilon_W, \varepsilon_{clay}$ are the bulk static dielectric constants of water and clay, and $\sigma_{eff}$ is the effective conductivity.

In the dielectric spectroscopy experiment the admittance as a function of frequency of the sample is measured from which the complex dielectric function $\varepsilon = \varepsilon' + i\varepsilon''$ is then deduced. The measured capacitance and conductance of the sample give the real and imaginary part of the dielectric function:

$$C(\omega) = \varepsilon' C_0, \quad G(\omega) = \varepsilon'' \omega C_0. \quad (7)$$

Here, $C_0$ is the capacitance of the empty measuring cell.
To compare the theoretical model for the conductivity of the clay-water system with the measured data, the conductance as a function of frequency is explicitly written:

$$G(\omega) = \text{Re}[\sigma(\omega)] \frac{C_0}{\varepsilon_0} = \frac{C_0 b}{\tau_b} \left( \frac{1}{1 + (\omega \tau_b)^2} + \text{sgn}(\alpha)(\omega \tau_s)^\alpha \right) + \sigma_{dc} \frac{C_0}{\varepsilon_0}. \quad (8)$$

Here, nondimensional parameter $b$ has been introduced with: $\varepsilon_0 b / \tau_b = \sigma_b$, where in case of diffusion controlled processes we have $\sigma_b = q^2 n D_{eff} / kT$ with $n$ representing the ion number density. The dependence on the sign of parameter $\alpha$ indicates that in case of superdiffusing particles the surface diffusion term conductivity may become negative leading to a situation where the energy flow in the system will increase instead of being dissipated. We must here emphasize two points: firstly the *total* conductivity of our system in the given frequency interval is always positive, and secondly, for a system under nonequilibrium conditions it is possible to exhibit negative conductivity, well known examples include certain kind of semiconductor and superconductor devices, and electrical networks with nonequilibrium elements. Negative conductivity has also been suggested for interacting classical Brownian particles and even single Brownian particle [15, 16].

The real and imaginary part of the dielectric function are not independent functions but are connected through Kramers-Kronig relations [20]:

$$\varepsilon'(\omega) = \varepsilon_\infty + \frac{2}{\pi} P \int_0^\infty \frac{x \varepsilon''(x)}{x^2 - \omega^2} dx. \quad (10)$$

Using this we compute the real part of the dielectric function from eqs. (8) of the model and compare it to the measured capacitance data using eqs. (9) and (7) . A large discrepancy between the calculated and measured capacitance would indicate to the inconsistencies in the proposed model for the conductivity (eq. (1)) which in our case was not observed.

## 4. Discussion

The results of the calculated conductance and capacitance using eqs. (7), (8) and (10) for the chosen set of parameters are presented on figs. 1-4 as solid lines. The values of the parameters obtained by fitting to the measured sets of data are given in table 1. The effective diffusion constant was calculated from eq. (4) using $a \approx 3/A_s \rho_w G_s$ [26] for the radius of the clay particle, where $A_s$ is the specific surface, $\rho_w$ is the density of water and $G_s$ is the specific gravity. The density of ions is then obtained from $\varepsilon_0 b / \tau_b = q^2 n D_{eff} / kT$ at room temperature.

The large increase of capacitance (or equivalently real part of the dielectric function) is often ascribed to electrode polarization effect in electrolyte solutions and colloidal suspensions. This is usually modelled with the equivalent circuit containing combination of frequency dependent capacitors and resistors with the high frequency $\omega^{-2}$ decay. A detailed analysis of the electrokinetic model for colloidal suspension [27] also yielded the same power law, $\omega^{-2}$ for the electrode polarization. On the other hand it was shown [28] that the real part of the dielectric function decays as $\omega^{-3/2}$ when the influence of the migrating charges in a finite space on the electric field is considered. Also it was found that the real part of the dielectric function starts to deviate from the high frequency value at crossover frequency $\omega_c \propto \sigma_\infty^{2/3}$ which scales with the high-frequency conductivity [28].

In our model presented here the high-frequency decay of the real part of the dielectric response follows $\varepsilon' \propto (\omega \tau_s)^{-(1-\alpha)}$ thus generalizing the model presented in [28], where the crossover frequency is in now given with $\omega_c = 1/\tau_s$. The (absolute) value of the parameter $\alpha$ increases with the water content in clay-water system reaching $\alpha = -0.51$ in the liquid limit. This observation suggests that the migrating charges influences the dielectric response of the clay-water system at liquid limit and at higher water content in the similar manner as in dilute colloidal suspensions, and that there is no principal difference between electrical properties of the saturated porous solids and colloidal suspensions as it was already suggested [29]. According to the high-frequency asymptotic form of the dielectric response, the scaling law for the crossover frequency for our model is $\omega_c \propto \sigma_\infty^{1/(1-\alpha)}$. The high-frequency value of the conductivity is expected to be also the function of water content, so in the lowest approximation we set $\sigma_\infty \approx |\alpha| \sigma_f$. With the help of eq. (4) we obtain in this approximation the scaling law for the effective diffusion constant $D_{eff} \propto (|\alpha| \sigma_f)^{1/(1-\alpha)}$. Using the values from table 1 the ratio of the obtained effective diffusion constants at the highest and lowest water content is 1,095 while the ratio of the corresponding expressions $(|\alpha| \sigma_f)^{1/(1-\alpha)}$ from the scaling law yields 1,086. Regarding the lowest approximation used this can be considered as a good agreement confirming the scaling of the effective diffusion constant. With the approximate expression for the conductivity of the pore electrolyte $\sigma_\infty \approx |\alpha| \sigma_f$ we can make a crude estimate of the dielectric constant of the clay-water electrolyte using eq. (5). With the pertinent values taken from table 1 we have $\varepsilon_w = 66$ at 56 % water content and $\varepsilon_w = 72$ at 32 % water content which roughly correspond to the values obtained in ref. [5] for the water in outer capillary and free water layer, again confirming the correct order of magnitude of the model parameters in table 1.

The parameters used in the model are not all independent. One constrained on the parameters follows from consideration of the frequency interval in which the negative real part of dielectric function is observed (see fig. 4). The boundaries of this interval follow from the solution of the equation $\varepsilon'(\omega) = 0$. Using eqs. (8) and (10) we rewrite this in nondimensional form:

$$c_1 + \frac{c_2}{x^{1-\alpha}} - \frac{1}{1+x^2} = 0 ,$$

where $c_1 = \varepsilon_\infty / b$, $c_2 = (\tau_s / \tau_b)^\alpha \tan(|\alpha|\pi/2)$, and $x = \omega \tau_b$. Fig 5. shows the real part of the measured dielectric response for the sample with 56 % moisture content normalized to minimum value (open dots) in logarithmic scale. Also shown are the plots of the normalized function $\varepsilon'(\omega)$, low frequency dispersion $c_1 + \frac{c_2}{x^{1-\alpha}}$, high frequency dispersion $\frac{1}{1+x^2}$, and asymptotic plots of the low and high frequency dispersion parts $c_1$ and $x^{-2}$ respectively. Form fig. 5 it is evident that in this case the solution of eq. (xx) for the lower and upper frequency boundary is to a good approximation given with $x_{low} \approx c_2^{\frac{1}{1-\alpha}}$ and $x_{high} \approx c_1^{-1/2}$. The ratio of the frequencies at the lower and upper boundaries in this approximation therefore imposes additional constraint on the parameters of the model.

Looking at results collected in table 1 we immediately notice very different values of parameters in cases of dry and moist clay. In dry clay the dielectric response of the sample origins from motion of ions near the clay particle surface in the bound water (adsorption and capillary) layer while in moist samples the free water electrolyte fills the space between the clay particles so the conductivity if the pore electrolyte adds to transport mechanisms. The change in the dynamics of transport mechanisms of ions in dry and moist samples is additionally signalled by the sign reversal of the parameter $\alpha$ from positive to negative respectively suggesting the crossover from sub- to superdiffusive regime with increasing water content. Somewhat strange values for the parameters in case of dry clay are a consequence of using the same model (e.g. eq. (8)) for full range of water content. The results in table 1 actually show that the frequency dependence of the conductivity of dry sample is simply given with $\sigma \propto \omega^\alpha$ with $\alpha = 0{,}47$ or that the only remaining term in the model for the conductivity (eq. (1)) is the due to the conductivity of the clay particle surface layer with possible bound water. This type of power law dependence of the conductivity on frequency is regularly observed in disordered systems [30,31], and also indicates to the fractal nature of the soil particle surface [schwartz-89,wong-86].

## 5. Conclusions

We have analyzed the dielectric response of the clay-water system at different levels of water content from dry to over liquid limit. The model for the frequency dependent conductivity is presented based on the dynamics of the ion in pore electrolyte, near the clay particle surface, and in free water. The parameters obtained from the fitting the model to the measured dielectric spectra are consistent with the expected values for clay-system. The dynamics of the ions near the clay particle surface is in this system found to be goverent by anomalous diffusion, and the analysis of the results lead to the approximate expressions for the scaling law of the effective diffusion constant. In case of dry clay the theoretical model yielded the power law frequency dependence of the conductivity typical for the disordered solids.

Table captions:

Table 1: Values of the parameters obtained by fitting the expression for the conductance and capacitance (eq. (9)) to the measured data and calculated effective diffusion constant and ions number density.

Tables:

Table 1

|  | $p=36\%$ | $p=56\%$ | dry clay |
|---|---|---|---|
| $\varepsilon_\infty$ | 2,5 | 2,5 | 2,5 |
| $\tau_b$ [s] | $1,8\times10^{-8}$ | $1,8\times10^{-8}$ | $2,9\times10^{-4}$ |
| $\tau_s$ [s] | $6,5\times10^{-4}$ | $5,8\times10^{-4}$ | $9,1\times10^{6}$ |
| $b$ | 950 | 1250 | $10^{-3}$ |
| $\sigma_f$ [S/m] | 0,18 | 0,13 | $-4,1\,10^{-7}\approx 0$ |
| $\alpha$ | -0,4 | -0,51 | 0,47 |
| $D_{eff}\times 10^{11}$ [m$^2$/s] | 2,1 | 2,3 | $1,5\times10^{-10}$ |
| $n\times 10^{-27}$ m$^{-3}$ | 3,7 | 4,3 | 3,4 |

Figure captions:

Figure 1: Measured conductance dependence on frequency for moist kaolinitic clay at 56 % (upper dots and curve) and 36 % (lower dots and curve) moisture content. The solid curves are calculated from the model described in the text.

Figure 2: Measured conductance dependence on frequency for dry kaolinitic clay. Empty dots represent measured values and the solid curve is calculated from the model described in the text.

Figure 3: Measured capacitance dependence on frequency for moist kaolinitic clay at 56 % (upper emtpy dots and curve) and 36 % (lower empty dots and curve) moisture content. The solid curves are calculated from the model described in the text

Figure 4: Measured capacitance dependence on frequency for moist kaolinitic clay at 56 % (upper emtpy dots and curve) and 36 % (lower empty dots and curve) moisture content. The frequency interval on this figure is chosen such to best demonstrate the observed negative capacitance effect. The solid curves are calculated from the model described in the text.

Figure 5: Analysis of the frequency interval in which the real part of the dielectric response of moist kaolinitic clay at 56 % moisture content exhibits negative values.

Figures:

Figure 1

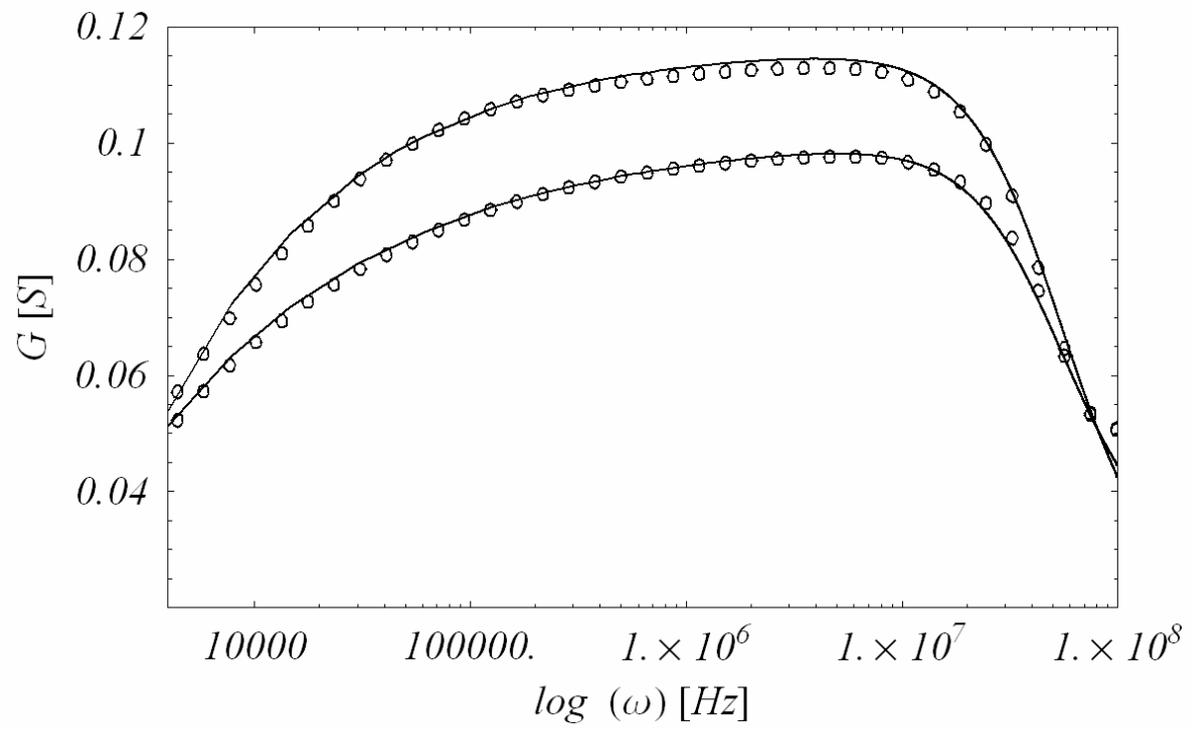

Figure 2

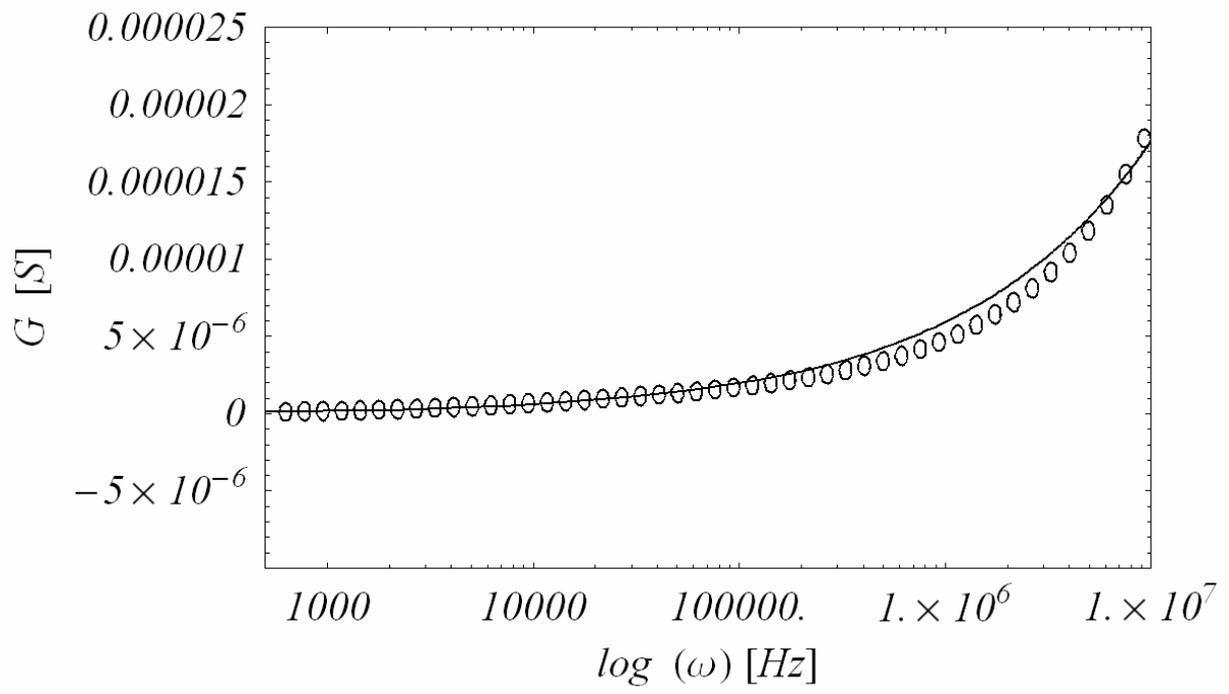

Figure 3

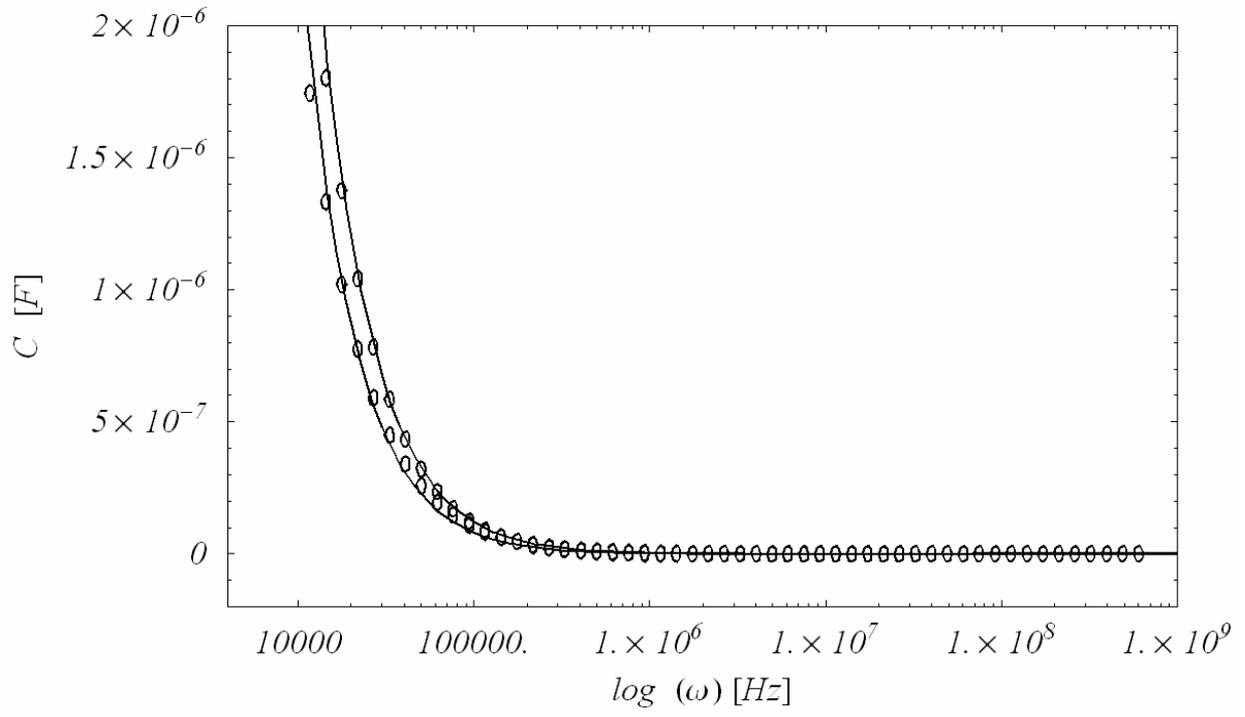

Figure 4

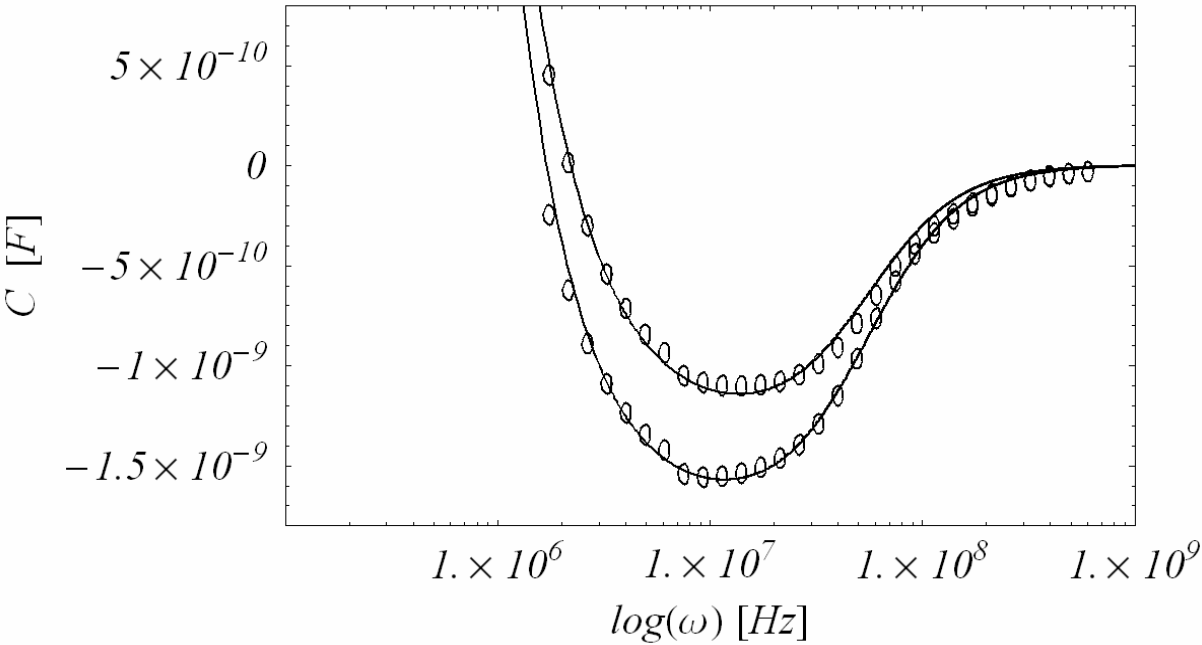

Figure 5

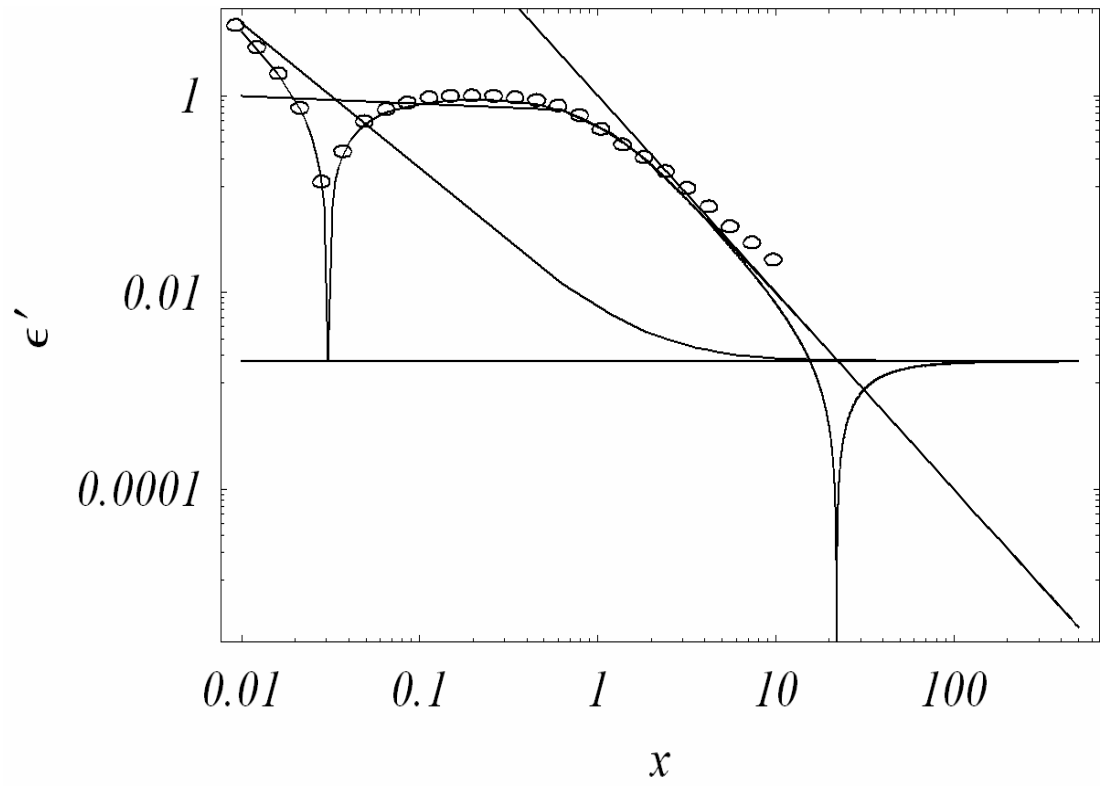